\documentstyle[emulateapj,psfig]{article}

\def\chandra{{\sl Chandra}}
\def\Chandra{{\sl Chandra}}

\def\simgt{\lower.5ex\hbox{$\; \buildrel > \over \sim \;$}}
\def\simlt{\lower.5ex\hbox{$\; \buildrel < \over \sim \;$}}
\newcommand\psr{\hbox{PSR~B$0833-45$}}

\slugcomment{To appear in {\it The Astrophysical Journal}}

\lefthead{Helfand, Gotthelf, Halpern}
\righthead{The Vela Pulsar X-ray Nebula}

\begin{document}

\title{The Vela Pulsar and its Synchrotron Nebula}

\author{D. J. Helfand, E. V. Gotthelf, \& J. P. Halpern}
\affil{Columbia Astrophysics Laboratory, Columbia University, 550 
West 120$^{th}$ Street, New York, NY 10027, USA}

\begin{abstract}

       We present high-resolution \Chandra\ X-ray observations of
\psr, the 89 ms pulsar associated with the Vela supernova remnant. We
have acquired two observations separated by one month to search for
changes in the pulsar and its environment following an
extreme glitch in its rotation frequency. We find a well-resolved
nebula with a toroidal morphology remarkably similar to that observed
in the Crab Nebula, along with an axial Crab-like jet. Between the two
observations, taken $\sim3\times10^5$s and $\sim3\times10^6$s after the
glitch, the flux from the pulsar is found to be steady to within $0.75
\%$; the $3 \sigma$ limit on the fractional increase in the pulsar's
X-ray flux is $\simlt 10^{-5}$ of the inferred glitch energy. We use
this limit to constrain parameters of glitch models and neutron star
structure. We do find a significant increase in the flux of the
nebula's outer arc; if associated with the glitch, the inferred
propagation velocity is $\simgt0.7c$, similar to that seen in the
brightening of the Crab Nebula wisps.

We propose an explanation for the X-ray structure of the Vela
synchrotron nebula based on a model originally developed for
the Crab Nebula.  In this model, the bright X-ray arcs
are the shocked termination of a relativistic
equatorial pulsar wind that is contained within the surrounding
kidney-bean shaped synchrotron nebula comprising the post-shock,
but still relativistic, flow.  In a departure from the Crab model,
the magnetization parameter $\sigma$ of the Vela pulsar wind
is allowed to be of order unity; this is consistent with the
simplest MHD transport of magnetic field from the pulsar to the
nebula, where $B \leq 4 \times 10^{-4}$~G.  The
inclination angle of the axis of the
equatorial torus with respect to the line of
sight is identical to that of the rotation axis of the pulsar as
previously measured from the polarization of the radio pulse.  The
projection of the rotation axis on the sky may also be close to the
direction of proper motion of the pulsar if previous radio
measurements were confused by orthogonal-mode polarized components.
We review effects that may enhance the probability of alignment
between the spin axis and space velocity of a pulsar, and speculate
that short-period, slowly moving pulsars are just the ones
best-suited to producing synchrotron nebulae with such aligned structures.
Previous interpretations of the compact Vela nebula as a bow-shock in
a very weakly magnetized wind suffered from data of inadequate spatial
resolution and less plausible physical assumptions.

\end{abstract}

\section {Introduction}
Within two years of the discovery of radio pulses from CP1919+21, magnetized,
rotating neutron stars were firmly established as the origin
of these remarkable signals. Furthermore, the steady increase in pulse
period recorded for all sources provided an explanation for the pulsar
power source: rotational kinetic energy. The detection of a decelerating
33 msec pulsar in the Crab Nebula solved the long-standing mystery of what
powered this unique nebula: the spin-down rate of the Crab pulsar
implied an energy loss rate, $\dot E \sim 5 \times 10^{38}$ erg s$^{-1}$,
more than enough to cover the radiation losses observed from radio to
gamma ray frequencies. By 1974, the basic model of the electrodynamics of
pulsar magnetospheres and their coupling to the surrounding synchrotron-emitting
plasma was in place (Rees and Gunn 1974), although a detailed understanding
of the processes involved continues to elude us (e.g., Arons 1998).

After a decade of timing observations, it became clear that most pulsars
were not defect-free clocks which simply slowed smoothly as rotational
energy was transformed into an electromagnetic outflow. Two distinct types of
non-monotonic behavior were established:
``timing noise'' characterized by a stochastic wandering in pulse phase and/or
frequency which appeared to afflict most pulsars (Helfand, Taylor, and Backus
1980; Cordes and Helfand 1980; Cordes and Downs 1985; D'Amico et al. 1998), and
``glitches'', an apparently instantaneous increase in the pulse frequency (a 
spin-up) accompanied by a simultaneous change in the spin-down rate; these rare
events were found to be most prevalent in young objects (Reichly and Downs 1971;
Lyne 1996 and references therein). Thirty years after the first glitch
in the Vela pulsar was recorded, a total of 65 events have been seen in 27
different pulsars (Lyne 1996; Wang et al. 2000).

Glitches are a sudden fractional increases in the pulsar spin frequency with
$\delta\nu / \nu \approx 10^{-9} \ {\rm to} \ 6
\times 10^{-6}$.  No pulsar with a characteristic age of $>10^6$ yr
has been observed to glitch more than once, but some young objects
experience these events roughly annually. The best studied and most
prolific in terms of large glitches is the first object in which a
glitch was seen -- the Vela pulsar. A dozen events have been recorded
over the past three decades and daily monitoring continues. The 
largest event yet observed occurred in 2000 January ($\delta\nu / \nu =
3.14 \times 10^{-6}$) and provided the stimulus for the observations
reported here.

In this paper, we report new results on \psr\ based on
observations acquired with the \Chandra\ High Resolution Camera.  The
data enable us for the first time to distinguish morphological details
of the synchrotron nebula surrounding \psr, and reveal a striking picture of
bilateral symmetry reminiscent of the loops and jets recently resolved
in the Crab Nebula (Weisskopf et al. 2000). We offer an interpretation
of the nebula's structure which requires an MHD wind with a high
magnetization parameter (unlike that inferred for the Crab).  We also
construct a high quality X-ray pulse profile and set tight upper
limits on any change in the profile following the glitch, constraining
models for the neutron star interior. Finally, we demonstrate an
apparent brightening in the Nebula a month after the spin-up event;
whether this was stimulated by the glitch or is a phenomena akin to
the flickering wisps in the Crab Nebula remains a question for future
observations to answer.

In section 2, we describe in some detail the analysis procedures
required to extract quantitative information from our HRC data, in
part as a cautionary tale for other early users of this instrument. We
then go on to explore the morphology of the pulsar's synchrotron
nebula (\S3), the soft X-ray pulse profile and limits on changes
thereto (\S4), and a search for changes in the nebula following the
glitch (\S5). The Discussion (\S6) begins by developing a model which
accounts for the nebula's geometry with respect to the pulsar, as well
as its anomalously low value of $L_X / \dot E$; we then go on to
derive constraints on glitch models from the temporal changes we see
(and don't see) following the glitch. The final section (\S7)
summarizes our conclusions and assesses the prospects for future
observations.

\section {Observations}

In response to an IAU Circular announcing a large Vela glitch on
2000 January 16.319 (Dodson et al. 2000), we submitted a Target of
Opportunity request to the \Chandra\ Observatory (Weisskopf, O'Dell, and van
Speybroeck 1996) to observe the pulsar as soon as practical, followed by a
second observation roughly one month later in order to search for
changes in the pulsar's flux, pulse profile, and/or surrounding
nebula. The observations were carried out on 20 January 2000 and 21
February 2000, $\sim 3.5$ and $\sim 35$ days after the glitch using
the  \Chandra\ imaging High Resolution Camera (HRC-I; Murray et
al. 1997). Integration times of $\sim 50$ ksec were achieved in both
observations.

The HRC-I detector on-board \chandra\ is sensitive to X-rays over the
$0.08-10.0$ keV range, although essentially no energy information on
the detected photons is available. Photons are time-tagged with a nominal
precision of 15.6
$\mu$s; in this work, their arrival times were corrected to
the solar system barycenter using a beta version of {\tt AXBARY}.  The
data were collected during a portion of the orbit which avoided
regions of high background contamination from the bright Earth
and radiation belt passages; the second observation was, however,
found to be partially contaminated by particle activity, most likely of
solar wind origin (see below). The pulsar was centered at the on-axis
position of the HRC where the point-spread function (PSF) has a
minimum half-power diameter (the radius enclosing 50\% of total source
counts) of $\sim 0\farcs5$, which increases with energy.  Images
were extracted centered on the pulsar and binned using
the native HRC $0\farcs13175 \times 0\farcs13175$ pixel size into
$1024 \times 1024$ pixel images ($2.5^{\prime}$ on a side).

We began our analysis using event data calibrated by the initial
processing and made available through the \Chandra\ public
archive. The first observation revealed several problems in the
standard data sets and further problems were subsequently found during
the analysis of the second observation. These problems affected both the
spatial and timing analyses and had significant implications for the
proper interpretation of the data. We alerted the HRC hardware and
software teams to the instrument and data processing anomalies, and
received considerable support in working through the problems. We
document here the various artifacts discovered and the steps taken to
correct for, or eliminate, them in our final data sets; our goals in
doing so are 1) to allow for the replication of our results, and 2) to
alert other early HRC users to problems they may encounter. In fact,
we found it necessary to reprocess the data from Level 0.5 using
custom scripts which incorporated improved filtering and processing
tools, making use of several beta versions of software provided by the
HRC team.

In the first observation we found evidence for a significant ``ghost
image'' which appeared as a spectacular jet-like feature emanating
from the pulsar along the detector $v$-axis. Examination by the
instrument team found that the standard processing had failed to screen
out all events flagged as instrumental. After filtering with a beta
version of {\tt SCREEN\_HRC} with the mask parameter set to $32771$, a
much truncated jet-like feature was still apparent. In order to
isolate any detector-centric artifacts, we obtained our second
observation with a roll angle offset by $36^{\circ}$ from the first. As
discussed below, we were able to confirm the reality of the residual
jet-like feature in the cleaned images.

Independent of any filtering, the initial images showed the pulsar to be
broader than the nominal PSF. To separate the pulsar from any proximate
nebular emission, we followed the same phase-resolved imaging analysis
described in Gotthelf \& Wang (2000) for their HRC observation of
PSR~0540$-$69, the 50 ms pulsar in the LMC. This separation, however,
failed completely.  By plotting the arrival times of the pulsar
centroid, we observed that the sky coordinates of the pulsar wandered
in a sinusoidal fashion with an amplitude of $0\farcs3$ and
the periodicity of the programmed telescope dither. This accounted for
the pulsar's non-point-like appearance in the time-integrated image.
Discussions with the \chandra\ attitude aspect team, however, showed a 
high-quality aspect reconstruction for the Vela observations.

Further analysis by the HRC team revealed a systematic problem with
one of the three anode preamplifiers which causes the coarse position
algorithm to mis-place photon locations depending on the photon input
position relative to the HRC tap gaps.  This explains the apparent
wandering of the pulsar centroid at the dither frequency: a fraction
of the detected photons are displaced along the detector coordinates
by a fixed amount. Indeed, we were able to ascertain
that the apparent diffuse flux was produced by faint echoes of the
pulsar itself along the two orthogonal detector axes
\footnote{ This problem does not affect early HRC-I observations obtained prior
to an increase in the instrument gain by a factor of two, such
as those of the 50 ms pulsar PSR~0540$-$69 (see Gotthelf \& Wang 2000).}.

To eliminate the echoes, we initially used the bright pulsar
as a fiducial point to re-aspect the photons and thus take out
the detector-induced wobble in a statistical sense. Subsequently, the
instrument team made available a beta version of a code to
identify and correct the mis-placed photons -- {\tt HRC\_EVT0\_CORRECT} --
along with updated degap parameters ($cfu1=1.068$; $cfu2=0.0$;
$cfv1=1.045$; $cfv2=0.0$) for use in {\tt HRC\_PROCESS\_EVENTS}.
This software, together with the new parameters, effectively
eliminated the echo problem. The images produced by the two methods are
indistinguishable and the pulsar now matches the PSF to within its
estimated uncertainty.

In the second observation, we noted additional artifacts in the
sky image resembling a rabbit-ear antenna extending
$20^{\prime\prime}$ from the pulsar in orthogonal directions along the
detector axes with point-like sources at the ends.  Temporal analysis
of the region showed that the rabbit-ear counts occurred during a
number of specific time intervals lasting tens of
seconds. Examination of the mission timeline parameters showed that
the occurrence of these events always followed the ``AOFF\_GAP'' times
by a few hundred seconds. Furthermore, data drop-outs were found for
tens of seconds at the ``AOFF\_GAP'' times and during the following
intervals when the spurious counts were recorded. We wrote an
algorithm to generate a new good-time-intervals file which eliminates
these intervals based on the ``AOFF\_GAP'' times.

Two additional detector issues needed consideration when extracting
accurate timing information: telemetry saturation and a hardware
time-stamp mis-assignment. Although data obtained during the first
observation displayed nominal background levels, the
second observation was plagued by intervals of telemetry saturation induced
by high background levels; such occurrences can seriously
affect timing studies by introducing 
spurious periods in the power spectrum aliased with the full buffer rate of
$\sim 4$ ms. We filtered out telemetry-saturated time intervals with the
dead-time fraction criteria of ${\rm DFT} > 0.9$. A further
complication for precision timing was recently discovered by the HRC hardware
team: the time-stamps for each event are mis-assigned to the following
event. Based on the {VALID\_EVT\_COUNT} count rate
of $\sim 500$ cps, the average error in the assigned photon arrival
time is 2 ms or a 2\% phase error for the Vela pulsar; assuming roughly
Poisson fluctuations in the HRC count rate over the observation interval,
the maximum error for any photon will be $\le 3$ ms. Thus, with 25 phase
bins across the 88 ms pulsar period, few, if any, photons have been 
misassigned and we have taken no mitigating action to correct this error.

To compare directly the two observations, we reprocessed both
data sets starting from the Level 0.5 event files using identical
methods and filter/screening/processing criteria, compensating for
incorrect keyword values, producing correct GTI extensions, etc. This 
resulted in a total of 50.3 ks and 45.3 ks integration times for the first and
second observation, respectively. Despite all the initial discrepancies and
artifacts in the two observations, this reprocessing produced effectively
identical images, light curves and count rates. We are thus confident that we
have eliminated all currently recognized instrumental artifacts in the
final data sets upon which we base the analysis herein.

\section {An Image of the Vela Pulsar}

A global view of the Vela pulsar and its environment as seen by the
\Chandra\ HRC is presented in Figure~1. The pulsar is embedded in a
complex region of previously resolved thermal X-ray emission from the
Vela supernova remnant that is present throughout this image and extends
far beyond its boundaries. The X-ray jet noted by Markwardt and Ogelman (1995)
is essentially overresolved in this image and extends far to the south of the
image boundary; it is evident as a faint enhancement in the diffuse
emission extending to the southeast and south of the bright pulsar nebula. 

The superb spatial resolution of the \Chandra\ HRC provides the first
look at the structure of the synchrotron nebula in the immediate
vicinity of \psr; Figure~2 shows an image constructed from the two
observations which have been centered on the pulsar and summed. The
bright point source representing the pulsar has an extent roughly
consistent with the local PSF. Apparently emanating from the pulsar,
towards the southeast, is a linear, jet-like feature
$10^{\prime\prime}$ in length.  There is also evidence for a counter
jet in the opposite direction. These jets have a position angle of
$130^{\circ}$ (measured East of North), and are aligned
to within $8^{\circ}\pm5^{\circ}$ degrees with the pulsar's proper motion vector
(Bailes et al. 1990; DeLuca, Mignani, and Caraveo 2000).

Concentric with the pulsar is a diffuse outer arc of emission
perpendicular to the jet. This feature is roughly
elliptical in shape and subtends an angle of $\sim 150$
degrees as seen from the pulsar. Interior to this arc is an elliptical ring of emission with a
curvature very similar to the outer arc. The pulsar, jet, and arcs are
embedded in a extended nebula of faint diffuse emission which has been described
as ''kidney-bean" shaped (Markwardt and \"Ogelman 1998). The
configuration of the jet feature relative to the nebula is reminiscent
of the \Chandra\ image of the Crab Nebula (Fig. 3; see Weisskopf et
al. 2000).

We determined the count rates by extracting counts from the various
regions discussed above (see Figure 4). For each source region we carefully estimated
the background.  For the diffuse emission, we determined the HRC
detector background derived from an annulus $13.2^{\prime\prime}$ wide 
exterior to the kidney bean emission ($r>52.7^{\prime\prime}$). We
extracted counts from the pulsar using a circular aperture 2\farcs64
in radius and estimated background from the surrounding annulus,
$2\farcs64 < r < 3\farcs43$. Table 1 summarizes the properties of the
individual components, including their estimated sizes and intensities.

\begin{deluxetable}{rcc}
\tablewidth{0pt}
\tablecaption{HRC Spatial Components of \psr
\label{tbl-1}}
\tablehead{
\colhead{}              & \multispan2{\hfil PSR~B0833--45~System \hfil} \\
\colhead{Component$^a$} &  \colhead{Shape and Size} &  \colhead{Counts s$^{-1}$$^b$} 
}
\startdata
A) Pulsar (Obs 1)& point-like & $2.012 \pm 0.006$  \nl
       (Obs 2)&            & $1.997 \pm 0.007$  \nl
B) Nebula (Obs 1)& $20^{\prime\prime} \times 10^{\prime\prime}$ NE-SW & $2.720 \pm 0.007$ \nl
 (Obs 2)&  & $2.762 \pm 0.008$  \nl
C) Arc  (Obs 1)& & $0.675 \pm 0.004$ \nl
        (Obs 2)& & $0.700 \pm 0.004$ \nl
D) Jet    (Obs 1)& $10^{\prime\prime}$ long SE-NW & $0.037 \pm 0.001$ \nl
    (Obs 2)& & $0.036 \pm 0.001$ \nl
\enddata
\tablenotetext{a}{See \S 3.}
\tablenotetext{b}{Count rates are background subtracted.}
\end{deluxetable}

The best current measurements for the Vela pulsar and remnant place it
at a distance of only $250\pm30$ pc (Cha, Sembach, and Danks 1999 and
references therein); in all that follows we scale by $d = 250 d_{250}$
pc.

Vela X, the $\sim 100^{\prime}$ diameter, flat-spectrum radio
component near the center of the Vela remnant (Milne 1968; Bock et al. 1998) is
generally regarded as the pulsar's radio synchrotron nebula. While
soft X-rays from this region are detected, they are primarily thermal
in nature, and represent emission from the hot plasma which fills the
entire remnant (Kahn et al. 1985; Lu and Aschenbach 2000). Even the bright radio
filament detected by Bietenholz, Frail and Hankins (1991) shows no
corresponding enhancement in our X-ray image, although more constraining
limits will be derivable from ACIS observations.

The compact X-ray source near the
pulsar was first recognized by Kellogg et al. (1973) as having a
harder spectrum. Subsequent observations with increasing angular
resolution (Harnden et al. 1985 and references therein; \"Ogelman,
Finley, and Zimmermann 1993; Markwart and \"Ogelman 1998) localized the
compact nebula to a region $\sim 2^{\prime}$ in extent roughly
centered on the pulsar. \"Ogelman et al. (1993) used the nominal PSF of
the ROSAT PSPC and an {\it ad hoc} model for the surface brightness of
the diffuse emission to attempt a deconvolution of the pulsar and its
nebula and to obtain spectral fits to the two components. They found
that a blackbody effective temperature of 0.15 keV adequately
characterized the point source, while the extended emission exhibited
a power law spectrum with a photon index of $\sim 2.0$; a column
density of $N_H = 1 \times 10^{20}$ cm$^{-2}$ is marginally consistent
with both components. Markwardt and \"Ogelman (1998) subsequently revised the
division of the flux between the point source and nebula based on
ROSAT HRI observations. Seward et al. (2000) attempted to isolate the
pulsar emission temporally and found somewhat lower blackbody
temperatures. The RXTE observations of Gurkan et al. (2000) found a
similar power law index for the nebula, but a much higher
normalization; while this could indicate diffuse synchrotron X-ray
emission from a larger area (given their one-degree field of view),
it could also result from background modeling problems, since Vela is
a weak source for RXTE.

As described above, our HRC image allows us to separate cleanly the
pulsar from the nebular emission. We find count rates for the pulsar
and the nebula minus the pulsar (within a radius of
$50^{\prime\prime}$) completely consistent with those of Markwardt and
\"Ogelman (1998) using the \"Ogelman et al.  (1993) spectral parameters
($kT = 0.15$ keV; $N_H = 1 \times 10^{20}$ cm$^{-2}$), confirming these
as a useful characterization of the observed flux.  This leads to an
unabsorbed, bolometric luminosity for the pulsar blackbody emission of
$\approx 1.5 \times 10^{32} d^{2}_{250}$ erg s$^{-1}$. Note, however, that
this luminosity is inconsistent with the adopted temperature ($T=1.7
\times 10^6$K) and distance (250 pc) for a uniformly radiating
blackbody with a radius of 10 km; even for the minimum neutron star
radius consistent with reasonable equations of state ($R\sim7$~km),
the derived $L_X$ is too high by a factor of 20. Lowering $T$ to $8.5
\times 10^5$K as advocated by Seward et al. (2000) yields
self-consistent values for $L_X, R, T$, and $d$, and raises the
intrinsic luminosity by $\sim30\%$.  Adjusting the distance, and
including such effects as non-grey opacity in the neutron star
atmosphere and a non-uniform temperature distribution over the surface
will also affect the calculated luminosity. For the purpose of a
comparison with glitch models (\S6), we adopt $T=1.0 \times 10^6$ K.

The integrated luminosity of the whole nebula
($r<52.7^{\prime\prime}$ minus the pulsar) in the $0.1-10$ keV band is
$3.5 \times 10^{32} d^{2}_{250}$ erg s$^{-1}$, corresponding to $4.9
\times 10^{-5}$ of the pulsar's spin down luminosity. This ratio of
$L_{neb}/ \dot E$ is significantly lower than that for any other
pulsar, and is a major constraint on models for coupling the pulsar
wind to the nebula (see \S6).

\section{The X-ray pulse profile}

After many unsuccessful searches, \"Ogelman et al. (1993) were the first
to detect X-ray pulsations from the Vela pulsar using the ROSAT
PSPC. Their observations revealed a complex profile, not obviously
related to the pulse profiles previously recorded at radio, optical,
and gamma-ray wavelengths. The observed pulsed fraction of $4.4 \pm
1.1\%$ was diluted by the inability of the PSPC to resolve the pulsar
from the surrounding nebula; using the approximate model described
above, the authors estimated a soft X-ray pulsed fraction of
11\%. Seward et al. (2000) constructed a higher signal-to-noise
profile by combining six ROSAT HRC observations; they estimated a
pulsed fraction of 12\% divided between a broad component (8\%) and
two narrow peaks (4\%). Strickman et al. (1999) and Gurkan et
al. (2000) have recently derived 2-30 keV profiles based on RXTE
observations of Vela. Strickman et al. illustrate a trend in which
the component separation of the main pulse increases with energy.

We have determined the X-ray pulse period for our two observations and
compared them to the radio ephemeris (Don Backer, personal
communication). We began by constructing a periodigram around a narrow
range of periods centered on the expected period $\pm 0.1$ ms, sampled
in increments of $0.05 \times P^2/T$, where $T$ is the observation
duration, and $P$ is the test period.  For each trial period, we
folded photons extracted from a $r = 2\farcs64$ aperture centered on
the pulsar position using 25 phase bins and computed the $\chi^2$ of
the resultant profile. We find a highly significant signal ($> 8
\sigma$) at $P = 89.32842(5)$ ms at epoch 51563.314043 MJD (TBD) and
$P = 89.32876(6)$ ms at epoch 51595.370251 MJD (TBD), completely
consistent with the observed radio periods. The uncertainty was
estimated according to the method of Leahy (1987). For each fold, we
adoped the period derivative determined from the radio ephemeris.  Our
measurements of the period are consistent (within the errors) with the
radio prediction.

In order to compare the pulse profiles for the two observations we used the
phase-connected radio ephemeris to fold and align them.  We
present the sum and difference profiles in Figure 5. This phase alignment
is completely consistent with one we computed
empirically by cross-correlating the two profiles; this suggests that
the \chandra\ clock is stable to a few
ms over a month. We can also derive the absolute
radio to X-ray phase offset if we assume that the absolute \chandra\
time assignment is accurate (the calibration of this quantity has not yet
been finalized). The phase of the radio peak relative to the X-ray profile
is indicated in Figure 5.

The $\sim 200,000$ counts, uncontaminated by nebular emission,
provides us with the highest signal-to-noise X-ray pulse profile for
Vela yet reported (Figure~5).  Greater than 99\% of the counts from
the blackbody component detected by the HRC fall in the $0.1-2.4$ keV
ROSAT band. We compute a pulsed fraction by integrating the counts in
the light curve above the lowest point and dividing by the total
counts within a radius of $r=2.64^{\prime\prime}$ and subtracting the
small amount of nebular background in this extraction radius (see
Table 1).  Our value for the pulsed fraction is $7.1 \pm 1.1\%$; the
quoted error is dominated by the Poisson uncertainty in the number of
counts recorded in the bin representing the light curve minimum.  This
value is somewhat lower than those cited above, but has the advantage
of utilizing a direct measure of the total point-source contribution
with subarcsecond resolution. We measure a separation between the two
peaks of the main component of $\delta\phi \sim 0.325$, consistent
with a linear extrapolation of the energy-dependence of this quantity
reported by Strickman et al. (1999).

The background-corrected pulsar count rates were found to be $2.012
\pm 0.006$ and $1.997 \pm 0.007$ c s$^{-1}$, respectively, for the first and
second epochs; the overall count-rate is constant to within 0.75\%
(the second observation is $1.5\sigma$ fainter than the first). Thus,
the $3\sigma$ limit on any increase in the pulsar luminosity in
response to energy input from the glitch is $<1.2 \times 10^{30}$ erg
s$^{-1}$ or $\Delta T \sim 0.2\%$, 35 days ($3 \times 10^6$ s) after
the event\footnote{This estimate includes a $\sim 20\%$ correction for
the fraction of the bolometric luminosity lying below the HRC
band.}. The lower half of Figure~5 shows the difference between the two observations
as a function of pulse phase, where the second dataset
has simply been scaled by the ratio of the total integration times; no
single bin has a discrepancy exceeding 1.5 sigma.  This constancy in
both the pulsed luminosity and pulse profile set interesting
constraints on the glitch mechanism (see \S6.4).

\section {Changes in the Nebula}

While the primary energy release from a glitch must be within (or on
the surface of) the neutron star, the response of the star's
magnetosphere could result in the release of energy to the
synchrotron nebula, triggering changes in its morphology and/or
brightness. Even without the stimulus of a glitch, the optical wisps
of the Crab Nebula near the pulsar have been shown to change on
timescales of weeks, presumably in response to instabilities in the relativistic
wind from the pulsar (Hester et al. 1995). Greiveldinger and Aschenbach (1999)
have also reported changes in the X-ray surface brightness of the Crab Nebula on
larger scales and somewhat longer timescales. Thus, we have examined our
two images of the Vela nebula carefully in a search for surface
brightness fluctuations.

We examined the count rate in the kidney-bean region bracketed by the
outer-arc and an inner circle with $r = 1\farcs32$ centered on the
pulsar. No significant change was observed between the two
observations. Similarly, no measurable change in the count rate
associated with the jet-like feature was found. Comparisons of other
regions defined in Table 1 also showed no change, with the notable
exception of the outer arc itself which appeared to increase in
brightness by $\sim 5\%$ in the second observation.

To investigate this further, we examined regions congruent with the
morphology of the nebula by constructing radial bins which are
elliptical in shape; the ratio of the semi-major to semi-minor axes of the
ellipse is 1.76, and the elliptical annuli are oriented at a
position angle of $50^{\circ}$ (east of north). As Figure~6 shows, the
sector of the radial profile encompassing the bright northwestern arc
exhibits a $7.8\sigma$ excess between semi-minor axis
radii of $13.5^{\prime\prime}$ and $18.0^{\prime\prime}$ in the sense
that the source is brighter in the second observation. No other
sectors or radii show any significant changes. In Figure 7, we display
the azimuthal profile of the whole nebula in the elliptical ring
$4.5^{\prime\prime}$ wide centered on these radii. The residuals are positive
(the second observation is brighter) throughout the range $250^\circ$
to $30^\circ$; the excess is significant at the $7.6 \sigma$ level
and represents a brightening of 5.3\%.  The excess energy being
radiated in the HRC band amounts to $\sim 3 \times 10^{30}$ erg
s$^{-1}$. For the assumed geometry discussed in $\S 6.1$, the outer arc lies at
a distance of $1.05 \times 10^{17}$ cm from the pulsar, requiring signal
propagation at $\simgt 0.7c$ if the impetus for the brightening
originated from the pulsar at the time of the glitch.

\section{Discussion}

\subsection{Geometry and Kinematics of the Nebular Structure}

Figure 8 shows our proposed model for the Vela X-ray nebula. We assume
that the two prominent arc-like features lie along circular rings
highlighting shocks in which the energy of an outflowing equatorial
wind is dissipated to become the source of synchrotron emission for
the compact nebula extending to the boundary of the ``bean''.  One
reason that the arcs are not complete rings might be that the
emission is from outflowing particles which Doppler boost their
emission in the forward direction.

This is essentially the picture for the similar arcs
surrounding the Crab pulsar first suggested by Aschenbach and Brinkmann (1975) and later elaborated by other authors (Arons et al. 1998 and references
therein). The main difference is that the dark cavity which
contains the unshocked pulsar wind in the Kennel \& Coroniti (1984a)
model of the Crab is small compared with the volume of the Crab
Nebula, while the radius of the Vela nebula (the bean), is barely
twice as large as its pulsar wind cavity (see Figure 4).  We assume that the two
rings straddle the equator symmetrically, and suppose that the deficit
of emission exactly in the equatorial plane is related to the fact
that this is where the direction of a toroidally wrapped magnetic
field changes sign; i.e., the field may vanish there.  The semimajor
axis of the ring $a = 25.7^{\prime\prime}$, and the ratio $a/b = 1.67$
specifies the angle that the axis of the torus (i.e., the rotation
axis of the pulsar) makes with the line of sight, $\zeta = {\rm
cos}^{-1}(b/a) = 53^{\circ}.2$.  The angle $\Psi_0 = 130^{\circ}$ is
the position angle of the axis of the torus on the plane of the sky,
defined according to convention as the angle measured to the east from
north.  The direction of rotation (the sign of $\Omega$) is arbitrary.

The projected separation of the two rings is measured as $s =
17.7^{\prime\prime}$.  The half opening angle of the wind $\theta$ is
then given by ${\rm tan}\,\theta\ =\ s/(2a\,{\rm sin}\,\zeta) = 0.43
(\theta = 23.\!^{\circ}3)$, and the radius of the shock is $r_s =
a\,d\,/{\rm cos}\,\theta$, where $d$ is the distance to the pulsar.
For $d = 250$~pc we find $r_s = 1.05 \times 10^{17}$~cm.

The rotating vector model of pulsar polarization (Radhakrishnan \&
Cooke 1969) is commonly used to derive information about the geometry
of the pulsar magnetic inclination and viewing angles.  The angles
$\zeta$ and $\Psi_0$ can in principle be evaluated independently using
information derived from polarization measurements of the radio pulse.
In particular, the swing in position angle $\Psi (t)$ of linear
polarization across the pulse is very sensitive to $\zeta$, the angle
between the line of sight and the rotation axis.  The angle $\alpha$
between the magnetic axis and the rotation axis is much more difficult
to measure unless $\alpha \approx \zeta$ -- {\it i.e.}, unless the line
of sight passes near the center of the polar cap.  Accordingly,
$\alpha$ is often assumed while $\zeta$ is fitted.  For example,
Krishnamohan \& Downs (1983) assume $\alpha = 60^{\circ}$ in their
model for Vela; this is consistent with the value $\alpha = 65^{\circ}$ derived by
Romani and Yadigaroglu (1995) from a fit of their geometric gamma-ray emission
models to Vela's pulse profile.  When an interpulse is observed, $\alpha$ is often
inferred to be $90^{\circ}$ (that is, both polar caps are visible in
this case).  With these definitions (see Figure (8) or Figure (13) of
Krishnamohan \& Downs (1983) for the geometry),

$${\rm tan}(\Psi (t) - \Psi_0)\ =\ {{\rm sin}\,\phi (t) \over
{\rm cot}\,\alpha\ {\rm sin}\,\zeta\ -\ {\rm cos}\,\zeta\
{\rm cos}\,\phi (t)}.\eqno(1)$$

\noindent
Here $\phi(t)$ is the longitude of the emitting region,
which increases linearly with time, and $\Psi_0$ is the
position angle of the rotation axis of the pulsar projected
on the sky as in Figure (8).

In the context of the rotating vector model, $\Psi_0$ is identical to
the position angle of polarization $\Psi$ at the peak of the pulse
where the magnetic dipole axis crosses the rotation axis ($\phi (t) = 0$
in Equation (1)).  Because the emission mechanism is thought to be
curvature radiation from particles moving along magnetic field
lines, the electric vector is tangent to those field lines, rather than
perpendicular to them as is the case with synchrotron radiation.  While
this measurement is in principle straightforward, in practice it is
not routinely accomplished.  Observations at two or more frequencies
are needed to determine (and to correct for) the interstellar rotation
measure, and to demonstrate that the intrinsic polarization is in fact
frequency-independent.  Another complication is that a pulse is often
composed of several identifiable components, some of which can be
polarized in the orthogonal mode (a result of propagation effects in
the magnetosphere) obscuring the ``true'' polarization.  Furthermore,
the pulse itself might not even contain an identifiable core
component, being composed instead of emission from random patches
within a cone (e.g., Lyne and Manchester 1988; Manchester 1995; Deshpande and Rankin 1999).  Accordingly, measurements
of $\Psi_0$ are rarely attempted, measurements of $\alpha$ are rarely
trusted, and measurements of $\zeta$ are rarely questioned.

In fact, there are several determinations of $\Psi_0$ for the Vela
pulsar that are not in particularly good agreement with each other; we
review a representative subset here.  The original value of
Radhakrishnan \& Cooke (1969) is $\Psi_0 = 47^{\circ}$ with an
uncertainty of $\approx 5^{\circ}$.  Hamilton et al. (1977) made
measurements over several years, all of which are consistent with
$\Psi_0 = 64^{\circ} \pm 1.\!^{\circ}5$.  A detailed decomposition
into four separate pulse components was performed by Krishnamohan \&
Downs (1983), in which they concluded that one of the components was
polarized in the mode orthogonal to the other three.  However, they
did not attempt an absolute measurement of the angle $\Psi_0$.
Bietenholz et al. (1991) measured $\Psi_0 = 35^{\circ}$
using the VLA (although this measurement may not be directly comparable, in
that it represents a mean value weighted by the degree of linear polarization
rather than the value at the center of symmetry of the position angle
curve).  Thus, while the published values of $\Psi_0$ differ by
as much as $30^{\circ}$, it appears that none is even close to being
aligned with the axis of the nebula, and that all are roughly
perpendicular to it.

Interestingly, the model of Krishnamohan \& Downs (1983) produces a
precise (albeit model-dependent) value of the angle between the rotation axis and
the line of sight, $\zeta = 55.\!^{\circ}57 \pm 0.\!^{\circ}15$.
This value agrees well with the inclination angle of our postulated
equatorial wind torus to the line of sight derived by fitting an ellipse to
the shape of the X-ray features: $\zeta = {\rm cos}^{-1}(b/a)
= 53{^\circ}.2$.  This striking coincidence gives us courage to
pursue the basic physics of the equatorial wind model using the
geometry of Figure 8, and even to be so bold as to suggest that
{\it all} of the radio determinations of $\Psi_0$ for Vela are
incorrect by $90^{\circ}$ because of incorrect mode identification
($\it i.e.$, perhaps three out of four of the pulse components are
actually polarized in the orthogonal mode).  In this case, $\Psi_0 =
130^{\circ}$ (as inferred from the orientation of the X-ray torus),
and can be identified with the projected direction of the pulsar
rotation axis.  Speculations about the true orientation of $\Psi_0$ in
pulsars go back to Tademaru (1977), who first discussed the possible
alignment between spin axis and proper motion in the context of the
radiation rocket hypothesis (Harrison \& Tademaru 1975).

\subsection{Implications of the Proper Motion}

For both the Crab and Vela pulsars, the direction of proper motion
(transverse velocity $v_t$) is strikingly close to the projected X-ray
symmetry axis of the inferred equatorial wind and polar jet-like
structures.  The proper motion of the Vela pulsar has been measured with
comparable accuracy using both radio interferometry (Bailes et al. 1990) and
images from
{\it HST} (De Luca et al. 2000).  The resulting mean value
$0.\!^{\prime\prime}056 \pm 0.\!^{\prime\prime}004$~yr$^{-1}$ at a
position angle of $302^{\circ} \pm 4^{\circ}$ is within $8^{\circ}$
of the axis of the nebula ($\Psi_0 + 180^{\circ} = 310^{\circ}$).  The
transverse velocity $v_t = 65$ km~s$^{-1}$ at $d = 250$~pc.  The
proper motion of the Crab pulsar is $0.\!^{\prime\prime}018 \pm
0.\!^{\prime\prime}003$~yr$^{-1}$ at a position angle of $292^{\circ}
\pm 10^{\circ}$ (Caraveo \& Mignani 1999), which corresponds to $v_t =
123$ km~s$^{-1}$ at $d = 2000$~pc.  The axis of the Crab's toroidal
optical and X-ray structure is $299^{\circ}$ (Hester et al. 1995),
only $7^{\circ}$ from the direction of proper motion.  The probability
that two such close alignments will occur by chance when drawn from a
pair of uncorrelated distributions is 0.7\%.  We also note that both
Vela and the Crab are rather slow moving compared to typical young
pulsars; Lyne \& Lorimer (1994) found a mean velocity for young pulsars of
between 400 and 500 km s$^{-1}$.

If these relationships are not a coincidence, then they may be
understandable in terms of the scenario proposed by Spruit \& Phinney
(1998) who suggested that the rotation axes and space velocities of
pulsars could be connected through the nature of the ``kicks'' given
to neutron stars at birth.  Spruit \& Phinney argue that the rotation
rate of the progenitor stellar core is too slow in the few years
before the formation of the neutron star for pulsar spin periods to be
explained by simple conservation of angular momentum during core
collapse.  Instead, it is likely that the same asymmetric kicks
(whatever the cause) that are responsible for the space velocities
of pulsars, are also the dominant contributors to their initial spin
rates.  If neutron stars acquire their velocities from a single momentum
impulse, then their rotation axes should be perpendicular to their
space velocities.  If, however, they receive many random, independently
located impulses over time, as might result from convection which
leads to anisotropic neutrino transport or anisotropic fallback,
then their velocities and spins should be uncorrelated in direction.
However, if those multiple thrusts are not short in duration relative
to the resulting rotation period, it is possible that kicks applied
perpendicular to the rotation axis will average out, while those that
are along the rotation axis will accumulate.  In the latter case,
particularly germane for short rotation periods, the space velocity
will be preferentially aligned with the rotation axis.  This is
actually the scenario preferred by Spruit \& Phinney, for which they
appeal to long duration (several-second) thrusts that could result
from the effect of parity violation in neutrino scattering in a
magnetic field.

Thus, in the context of the above models, we speculate that the Crab and Vela
pulsars have relatively
low space velocities because the components of their kicks
perpendicular to their rotation axes were averaged out and, as a
result, their final space velocities were aligned closely with their spin axes
(cf. Lai, Chernoff, and Cordes 2001). We also note that alignment of the spin
axis and proper motion is a natural consequence of the Harrison and Tademaru 
(1975) photon rocket acceleration mechanism. With the recent revision of Lai et 
al. (2001), an initial spin period as long as 6 ms suffices to account for the
measured transverse component of the velocity in the maximum acceleration case.
Since pulsars that rotate most rapidly at birth are also the
ones most capable of powering synchrotron nebulae, either scenario
might argue for a stronger than average correlation of the axes
of such nebulae with the proper motion directions of their parent pulsars.

Before the toroidal arcs in the Vela nebula were resolved by
{\it Chandra}, Markwardt \& \"Ogleman (1998) interpreted the overall
shape of the nebula as seen by the {\it ROSAT\/} HRI as being {\it determined}
by the space velocity of the pulsar.  In particular, they noted that the
outline of the nebula, which is dominated by the bean shape,
resembles a bow shock whose symmetry axis at position angle
$295^{\circ}$ is identical to the direction of the radio proper motion
($297^{\circ}$).  However, the rather uniform and gently curved
outline of this structure could only be reconciled with the sharper,
asymmetric curve expected of a bow shock if the space velocity were
nearly along the line of sight.  This requirement, coupled with
the large absolute velocity
needed for the pulsar to exceed the speed of sound in the surrounding
supernova remnant forced Markwardt \& \"Ogleman to conclude 
that the velocity vector is less than $22^{\circ}$
from the line of sight.
This notion of the compact Vela nebula as a bow
shock also led Chevalier (2000) to a considerably different model of 
its physics.  As we shall argue below, the {\it Chandra} observations do not
support such a bow-shock interpretation, but instead favor a physical
model in which the entire structure is a synchrotron nebula
similar in physics to the Crab, but with an interesting difference
in one of its parameters.

\subsection{A Physical Model of the Nebula}

In the basic Kennel \& Coroniti (1984a,b) model of the Crab Nebula,
a relativistic pulsar wind terminates in an MHD shock, which produces
the nonthermal distribution of particles and post-shock magnetic field
that comprise the synchrotron nebula.  Although the pulsar wind
is assumed to carry the entire spin-down luminosity of the pulsar,
specific wind parameters such as the particle velocity and the fraction
of the power carried in magnetic fields are not known
{\it a priori}.  Rather, they are inferred by using the
results of the shock jump conditions to model the spectrum of the Crab
Nebula, and also to match the observed radii of the MHD shock and the 
outer boundary of the Nebula.  It is necessary to adopt
outer boundary conditions; a natural one is to require the final velocity
of the flow to match the observed expansion velocity of the Nebula,
although it is not clear how this outer boundary condition is communicated
to the inner MHD shock which is a factor of 20 smaller in radius.

A peculiar result of the Kennel \& Coroniti model is that the wind
magnetization parameter $\sigma$ is required to be $\approx 0.003$
in the Crab.  That is, the fraction of power
carried in $B$ field is much less than 1\%.  Such a small fraction is
required in order that sufficient compression occurs in the shock to
convert the bulk flow energy into random energy of the particles
so that they can radiate the observed synchrotron luminosity.
Highly magnetized shocks produce less radiation because there
is little energy dissipation and, for the Crab, would supply an
insufficient number of X-ray emitting particles.  Furthermore, highly magnetized
shocks are weak because all of the
energy dissipation allowed by the jump conditions is used in making the
small increase in $B$ field needed to conserve magnetic flux.
The post-shock flow velocity is still relativistic.

The reason that such a small magnetization is difficult to understand
is that the pulsar magnetic field energy carried out to the radius of the
shock in the simplest MHD wind should be of the same order of magnitude 
as the spin-down power, as the following argument shows.
The total wind energy flux at the shock is

$${I\,\Omega\,\dot\Omega \over 4\pi\,r_s^2}\ =\
\left ( {B_p^2\,R^6\,\Omega^4 \over 6\,c^3}\right )
\ {1 \over 4\pi\,r_s^2}\ =\ 
{B_p^2\,R^6\,\Omega^4 \over 24\pi\,c^3\,r_s^2}
\eqno(2)$$

\noindent
while the transported pulsar magnetic field $B_s$ at the
location of the shock is

$$B_s \approx {B_p  \over 2} \,\left ( {R \over r_{lc}} \right )^3\,
\left ( {r_{lc} \over r_s} \right )\ =\ 
{B_p\over 2} {R^3\,\Omega^2 \over c^2\,r_s}\eqno(3)$$

\noindent
where $r_{lc}$ is the radius of the light cylinder defined as $r_{lc} = 
c/\Omega$ and $R$ is the neutron star radius at which the
magnetic field strength is $B_p$.  Therefore, the magnetic energy flux at $r_s$ is

$$c\,{B_s^2 \over 8\pi}\ \simeq\ {B_p^2\,R^6\,\Omega^4 \over 32\pi\,c^3\,r_s^2}\ ,
\eqno(4)$$

\noindent
comparable to the value in Equation (2).
While various solutions to this paradox for the Crab have been
proposed, we argue here that the dimensions and spectrum of the Vela
synchrotron nebula are in much better accord with $\sigma \sim 1$.

A basic application of the pulsar wind model to the Vela synchrotron
nebula was made by de~Jager, Harding, \& Strickman (1996) in conjunction
with their detection of Vela with the Oriented Scintillation
Spectrometer Experiment (OSSE) on the {\it Compton Gamma-Ray Observatory}.
We summarize their conclusions here.  de~Jager et al.
noted that the unpulsed part of the hard X-ray spectrum extends
with power-law photon index $\Gamma = 1.73$ up to 0.4~MeV, which implies
that the synchrotron nebula radiates
$\sim 2 \times 10^{33}(E_{\rm max}/0.4\,{\rm MeV})^{0.27}$ ergs~s$^{-1}$,
or only $\sim 3 \times 10^{-3}$ of the pulsar spin-down power.
The absence of an observed spectral break limits the residence time $\tau_r$
of the electrons in the nebula that radiate in this energy range
to less than their synchrotron lifetime $\tau_s = 5.1 \times 10^8/
(\gamma\,B^2)$~s.  Since the electrons radiating at the highest observed
energy $E_{\rm max}$ have $\gamma = [2\pi\,m\,c\,E_{\rm max}/(h\,e\,B)]^{1/2}$,
the upper limit on the nebular magnetic field is $B < 20\ \tau_r^{-2/3}$~G.
To make use of this limit,
de~Jager et al. assumed that the residence time would be $\tau_r \approx r/c_s$
where the nebula outer radius $r \sim 2 \times 10^{17}$~cm, and $c_s = c/\sqrt 3$, the velocity
of sound in a relativistic plasma.  For these values,
$B < 4 \times 10^{-4}$~G.  This requirement is compatible with
a strongly magnetized pulsar wind for which little change in $B$
occurs across the shock, while the post-shock field continues to decline
inversely proportional to the distance from the pulsar.  Indeed, equation (3) predicts a pre-shock field of
$\sim 1.5 \times 10^{-4}$~G, consistent with the limit derived above; thus,
the wind may remain relativistic
across the nebula, which extends a factor of 2 in radius
beyond the shock. 

Such a synchrotron nebula is in approximate pressure balance
with the surrounding supernova remnant.  Markwart \& \"Ogelman (1997)
found by fitting a two-temperature thermal model to the {\it ASCA\/} spectrum
of the inner remnant that the thermal pressure is $\approx 8.5 \times
10^{-10}$ erg~cm$^{-3}$.  This compares well with the pulsar wind pressure
at $r$, $\dot E/(4\pi\,r^2\,c) = 5.2 \times 10^{-10}$ erg~cm$^{-3}$.
The entire compact X-ray nebula, then, is consistent with being powered
by a strongly magnetized pulsar wind shock whose still relativistic
downstream flow is confined by the Vela SNR.  The modest production of
synchrotron electrons in such a shock naturally explains the very low value
of $L_X/\dot E$ observed. 

In summary, the match between the nebular field upper limit derived from
the {\it OSSE} data and the value found from equation 3 using our
measured value of $r_s$, and the fact that the pressure corresponding to
this field strength matches the confining thermal pressure of the X-ray gas
leads us to conclude that our high-magnetization model of the Vela nebula
is both self-consistent and plausible. Thus, in a sense, Vela may be a more
natural realization of the Kennel \& Coriniti model than is the Crab,
for which the model was created.
 
While this appears to be a quite satisfactory model, certain details
are subject to additional constraints.
First, since the radiation from the
nebula is so inefficient, energetic electrons must be able to
escape to much larger distance scales before losing all of their energy to
synchrotron radiation. Indeed, the radio luminosity of the 100-arcminute
Vela X region,
$\sim8 \times 10^{32}$ erg s$^{-1}$, could be one manifestation of the escaping
electrons.  Second, there is a natural upper energy to the
synchrotron spectrum when the electron gyroradius $r_g$
exceeds the radius of the nebula.   Since
$$r_g\ =\ 1.6 \times 10^{17}\
\left ( {B \over 10^{-4}\ {\rm G}} \right )^{-3/2} \ 
\left ( {E \over 100\ {\rm MeV}} \right )^{1/2} \ {\rm cm}
\eqno(5)$$
this is not a restrictive limit.  Thus, we find the de~Jager et al (1996)
description of the Vela synchrotron nebula basically in accord with the
{\it Chandra} observations.

An alternative picture was proposed by Chevalier (2000) based on the
bow-shock interpretation of Markwardt \& \"Ogleman (1998), and a
simplified version of the Kennel \& Coroniti model.  If the Vela
synchrotron nebula is energized by a shock between a relativistic
pulsar wind and the surrounding supernova remnant, then the bow shock
travels at the velocity of the pulsar $v_p$, which necessarily exceeds
the sound speed in the hot confining medium.  In the first place, such
a large pulsar velocity is hardly likely, since the thermal sound
speed $\sqrt{dP/d\rho}\,\approx\,875$ km~s$^{-1}$ in the Vela SNR
according to the {\it ASCA\/} spectral analysis of Markwardt \&
\"Ogleman (1997).  Markwardt \& \"Ogleman (1998) assumed that $v_p
\geq 260$ km~s$^{-1}$ is sufficient.

An additional consequence of this scenario, however, is that the
residence time of the emitting particles in the nebula is much longer than
in the de~Jager et al. model, $\tau_r \sim r/v_p$ instead of $\tau_r
\sim \sqrt 3 r/c$.  Consequently, Chevalier (2000) was forced to
assume $\tau_r \sim 10^3$~yr, requiring an extremely small
magnetization parameter, $\sigma < 10^{-4}$, in order that the model's
radiated luminosity not exceed the observed X-ray luminosity. We consider that 
the physical difficulties of the bow-shock
interpretation, in conjunction with the new {\it Chandra} evidence
that the X-ray morphology is dominated by a pair of arcs resembling
similar toroidal structures in the Crab Nebula,
strongly disfavor such a model. Chevalier's revised estimate of $\sigma$ is
0.06 after adopting our assumption of relativistic post-shock flow. Determining
whether $\sigma$ is actually of order unity or not will require more detailed 
modeling

While our model provides a plausible explanation for the torroidal arcs and 
outer nebula, it says nothing about the other striking feature of the image,
the jet and counterjet. The jet is $10^{\prime\prime}$ long, giving it a
deprojected length of $4.1\times 10^{16}$ cm. At $v=65$ km s$^{-1}$, it
has taken the pulsar $\sim 200$ yr to travel this distance. The synchrotron
lifetime of electrons producing 1 keV emission ranges from 5 yr to 40 yr
for the fields of $(1-4)\times 10^{-4}$ G discussed above. Thus, the jet
is not simply a wake, but must be supplied with particles from the pulsar
(as the existence of the counterjet also suggests). The luminosity required
is modest: $L_X \sim 5 \times 10^{30}$ erg s$^{-1}$, roughly 1\% of the
nebular luminosity and $< 10^{-6} \dot E$. While we have no scenario
to propose, the existence of such features in both Vela and the Crab
suggests understanding their origin may prove useful in
modeling particle flow from young pulsars.

\subsection{Thermal Emission Constraints on the Neutron Star Interior}

Several models have been advanced to explain the sudden apparent
change in the moment of inertia of a glitching neutron
star. Originally, starquakes, resulting from the release of strain in
the stellar crust induced by the change in the equilibrium ellipticity
of the star as it slows, were invoked (Ruderman 1969). But the
magnitude and frequency of the Vela glitches could not be explained by
this model, and a picture involving the sudden unpinning from the
inner crust of superfluid vortices in the core of the star became the
dominant paradigm (Anderson and Itoh 1975). Observations of the
relaxation of the star back toward its original spin-down rate suggest
that $\sim 1\%$ of the star's mass is involved in the event (Alpar et
al. 1988; Ruderman, Zhu, and Chen 1998), implying a total energy
release of $\sim 10^{42}$ ergs.

The fate of this energy is unclear and predictions concerning the
observable consequences vary widely. The timescale for energy
deposited at the base of the crust to diffuse outward, the fraction of
the surface area whose temperature will be affected, and the secondary
effects, such as the rearrangement of the surface magnetic field which
could dump energy into the surrounding synchrotron nebula, are all
uncertain by one or more orders of magnitude. 

Our stringent upper limit on a change in the X-ray flux from the
neutron star within the 35 days following the glitch allows us to begin
setting meaningful constraints on the parameters of the neutron
star and the glitch.  Seward et al. (2000) provide a concise
introduction to the published models for the thermal response of the
stellar surface to a glitch generated by the sudden unpinning of
superfluid vortex lines deep in the star (Van Riper, Epstein, and Miller 1991;
Chong and Cheng 1994; Hirano et al. 1997, and Cheng, Li, and Suen
1998), and we need not repeat it here. We follow their approach in
deriving parameter limits from the models.

The allowable change in the pulsar's flux derived in \S4 corresponds to
a fractional change in the surface temperature of $ <0.2\% $. For
timescales of $\sim 30$ days, we are primarily sensitive to stars with
small radii ($R<14$~km) which correspond to soft or moderate equations
of state. Using Figure 2 of Van Riper et al. (1991), we can set a
limit of $E_{glitch} < 10^{42}$ ergs independent of the depth of
occurrence within the inner crust. For depths corresponding to local
densities $ \rho < 10^{13}$ gm cm$^{-3}$, $E_{glitch} < 3 \times
10^{41}$ ergs, and for shallow events ($\rho \sim 10^{12}$ gm
cm$^{-3}$), the glitch energy must be less than $10^{41}$ ergs. For
the softest equation of state used by Hirano et al. (1997)
corresponding to a $1.4M_{\odot}$ star with a radius of 11 km, glitch
depths shallower than $10^{13}$ gm cm$^{-3}$ require an energy
deposition of less than $10^{41}$ ergs. With observations of similar
sensitivity $\sim300$ and $\sim3000$ days after the event, we could
rule out $E_{glitch} \sim 10^{43}$ erg for all equations-of-state and
glitch-depth combinations, and require $E_{glitch}<10^{41}$ ergs for
soft and moderate equations of state for depths $\rho <10^{13.5}$ gm
cm$^{-3}$. Note that on the longest timescales (appropriate for deep
glitches in stars with very stiff equations of state), Vela may be an
inappropriate target for constraining glitch parameters, since another
glitch may well have occurred before the thermal pulse has peaked;
indeed, if even 10\% of the glitch energy appears as surface thermal
emission, the total X-ray luminosity can be powered by events with
$<E_{glitch}>\sim2 \times 10^{41}$ ergs.

\section {Summary and Conclusions}

We have presented a high-resolution X-ray image of the Vela pulsar
revealing a highly structured surrounding nebula. We interpret the
nebula's morphology in the context of the shocked MHD wind model
developed by Kennel and Coroniti (1984a,b) for the Crab Nebula, and find
that the Vela nebula allows a large magnetization parameter,
possibly of order unity. This
picture also provides a natural explanation for the low $L_X/\dot E$
of the Vela nebula. We speculate that the alignments of the symmetry
axes of the Crab and Vela nebulae with the proper motion vectors of
their respective pulsars should be expected preferentially in rapidly
spinning young pulsars with surrounding X-ray synchrotron nebulae if
the causal connection between spin and proper motion suggested by Spruit
and Phinney (1998) is correct.

Our two observations, centered 3.5 and 35 days after the largest
glitch yet recorded from the pulsar, allow us to set significant
limits on changes in the pulse profile and stellar luminosity which
can be used to constrain glitch model parameters. We find that, for
soft and moderate equations of state, the glitch energy must be
$<10^{42}$ ergs; an additional observation a year following the event
will substantially tighten this constraint.  An apparent change in the
nebula surface brightness between the two observations may or may not
be a consequence of the glitch; the implied velocity of the
disturbance, assuming that it originates near the pulsar, is $\sim 0.7
c$, similar to the velocity inferred from changes in the Crab Nebula
wisps.

Future observations with \chandra\ and XMM can be used to gain further
insight into the structure of the neutron star and its surrounding
nebula. An observation $\sim 1.5$ yr after the glitch, now scheduled,
 will further
constrain models for the glitch and the parameters of the neutron
star.  Additional HRC observations will also be required to decide
whether the nebula changes reported here are a consequence of the
glitch, or whether they occur routinely in response to instabilities
in the pulsar's relativistic wind, as appears to be the case in the
Crab Nebula. Data from the EPIC PN camera on XMM will yield spectral
clues helful in understanding the complex pulse profile, while either
EPIC or \chandra's\ ACIS could be used to search for spectral changes
caused by synchrotron energy losses and/or internal shocks in the
nebula.

\acknowledgements

We are grateful to the Chandra Science Center Director, Dr. Harvey
Tananbaum, for making this TOO possible. We also wish to acknowledge
Dr. Steven Murray and Dr. Michael Juda for many extremely helpful
discussions concerning HRC issues, and for kindly making available
beta-version software. We thank Drs. Fernando Camilo and George Pavlov
for discussion of
the pulsar timing issues and assistance in determining the radio pulse
phase, and Don Backer for providing the radio ephemeris. This work was
funded in part by NASA LTSA grant NAG5-7935 (E.V.G.), and SAO
\Chandra\ grant DD0-1002X (D.J.H).  This is contribution \#692 of the
Columbia Astrophysics Laboratory.


\begin{figure} 
\centerline{ {\hfil\hfil
\psfig{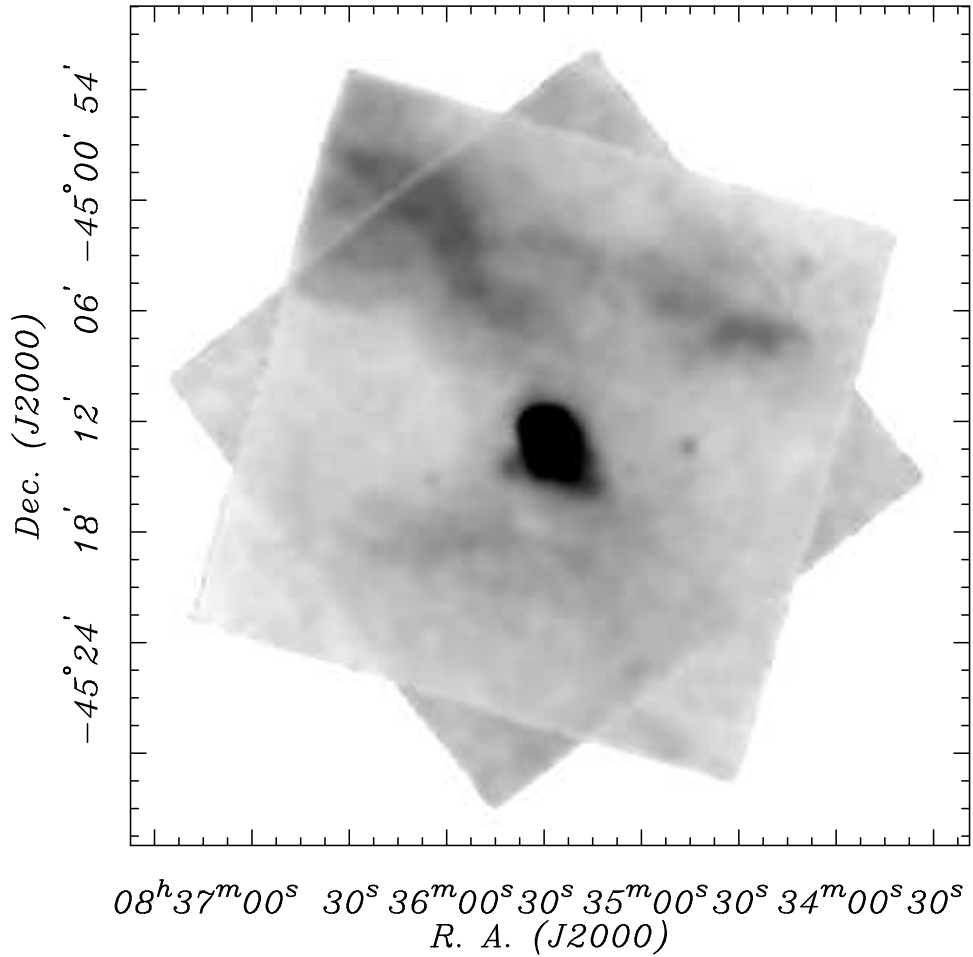}
\hfil\hfil} }
\caption{An X-ray image of the region containing the
Vela pulsar \psr. The two \Chandra\ High Resolution Camera (HRC) X-ray
datasets (0.1-10 keV) centered on the pulsar were summed and smoothed with a
Gaussian with $\sigma=0.5^{\prime}$. The intensity scale is chosen to highlight the
diffuse SNR emission surrounding the bright pulsar nebular which is
fully saturated in this image. }
\end{figure}

\begin{figure} 
\centerline{ {\hfil\hfil
\psfig{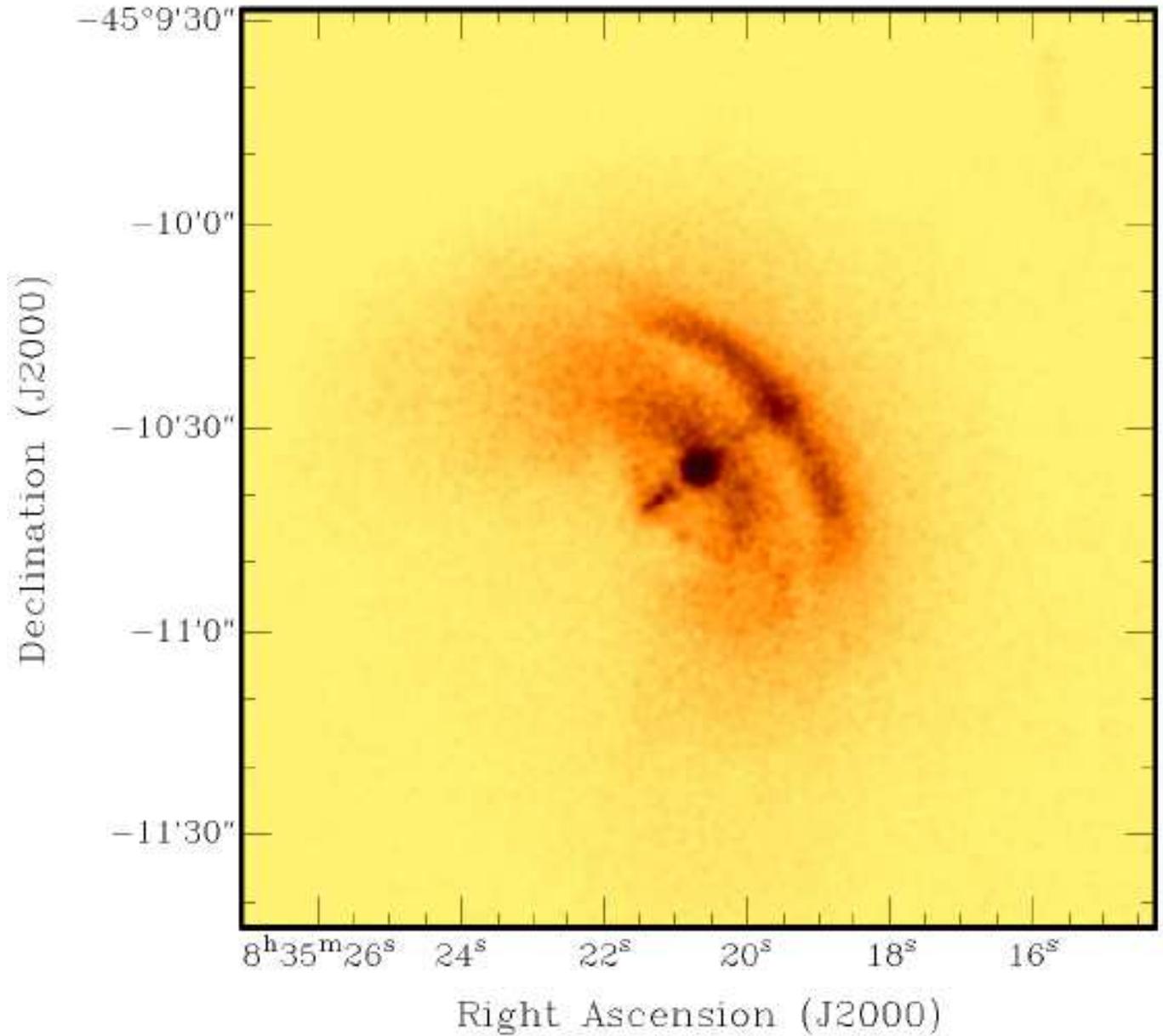}
\hfil\hfil} }
\caption{A close-up view of the region surrounding the Vela pulsar
\psr. The image includes the summed data from two epochs separated by a
month and is centered on the pulsar and scaled to highlight the
surrounding nebula emission. The data in this and the two subsequent
figures have been smoothed with a Gaussian with
$\sigma = 0.66^{\prime\prime}$.
A toroidal structure and
perpendicular jet similar to that seen in the Crab Nebula is apparent. Also
evident is a faint halo of emission likely associated with the
post-shock pulsar wind (see text).
}
\end{figure}

\begin{figure} 
\centerline{ {\hfil\hfil
\psfig{figure=vela_hgh_f3a.ps,height=3.0in,angle=270, clip=}
\quad
\psfig{figure=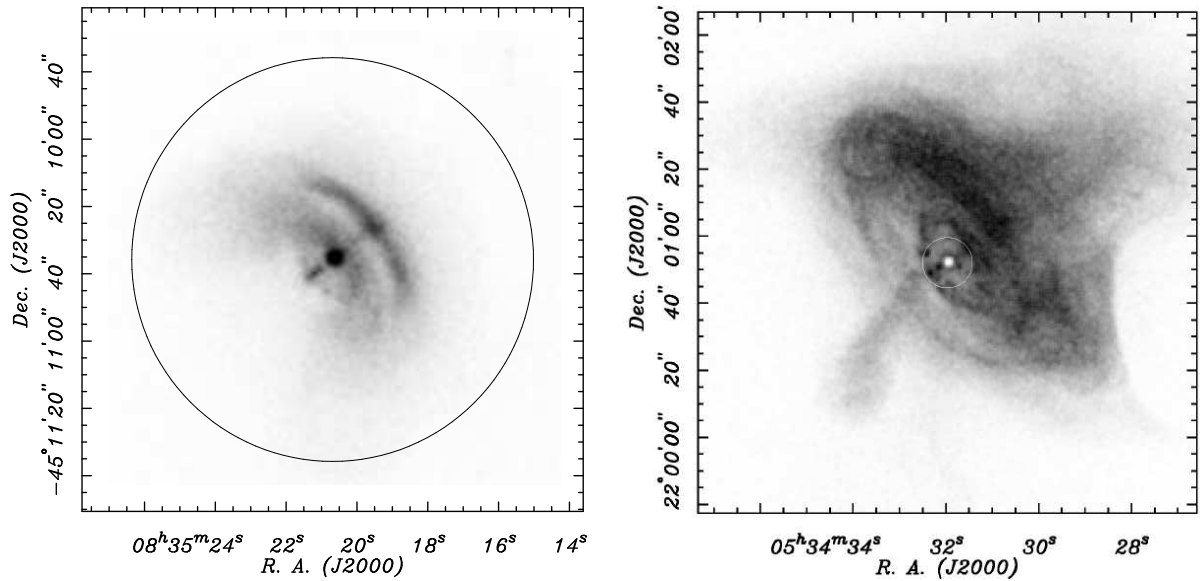,height=3.0in,angle=270, clip=}
\hfil\hfil} }
\caption{A comparison of the relativistic wind nebulae surrounding two
young pulsars observed by the Chandra Observatory: the 1000-yr Crab
pulsar (right) and the 10 kyr Vela pulsar (left). The images are
displayed with the same plate scale, although the Vela nebula is
sixteen times smaller physically assuming distances of 2 kpc (Crab)
and 250 pc (Vela); the circles in the two images represent the same
physical size at the respective pulsars.}
\end{figure}

\begin{figure} 
\centerline{ {\hfil\hfil
\psfig{figure=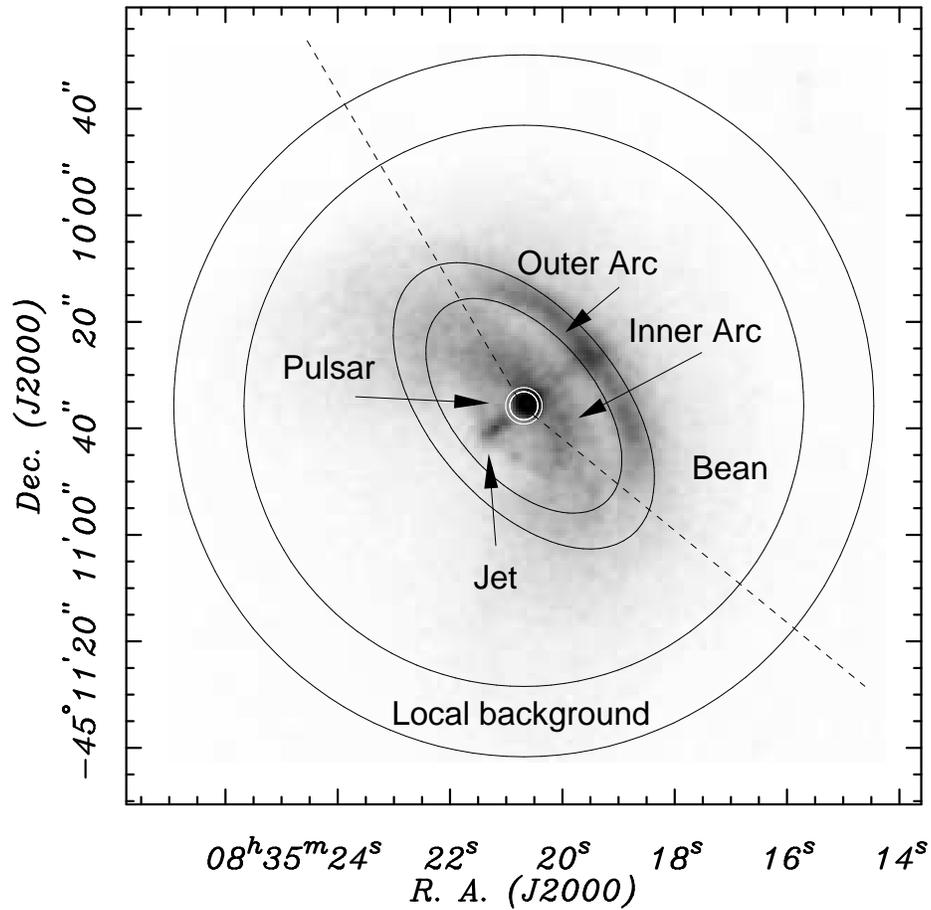,height=6.5in,angle=270,clip=}
\hfil\hfil} }
\caption{ The same image as in Fig. 2 with the various extraction
regions described in the text shown: the circular extraction and
background anulii region for the pulsar, the elliptical annulii
bounding the toroidal structure which define the region used for the
azimuthal profiles, the sector used to extract the radial profile (dashed
lines),  and finally, the
background anulii (see text for details). The sector bounded by the 
the elliptical annulii and dashed lines is found to
brighten between the two observations. }
\end{figure}

\begin{figure} 
\centerline{ {\hfil\hfil
\psfig{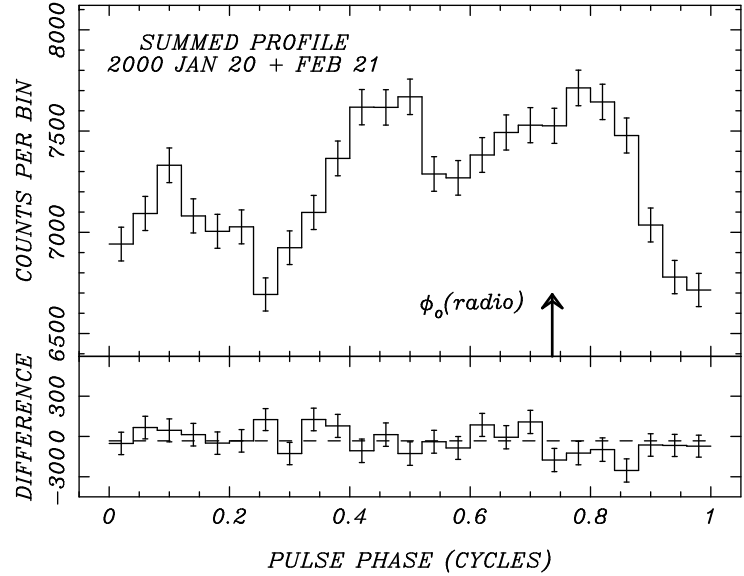}
\hfil\hfil} }
\caption{Top panel: the X-ray ($0.1 - 10$ keV) pulse profile of the Vela pulsar
obtained with the \chandra\ HRC from the combined observations. Bottom panel:
the difference profile -- 2000 21 February minus 2000 20
January.  There is no significant pulse shape change at
any phase. The phase $\phi_o$(radio) is the location of the radio
pulse.
}
\end{figure}

\begin{figure}
\centerline{
\psfig{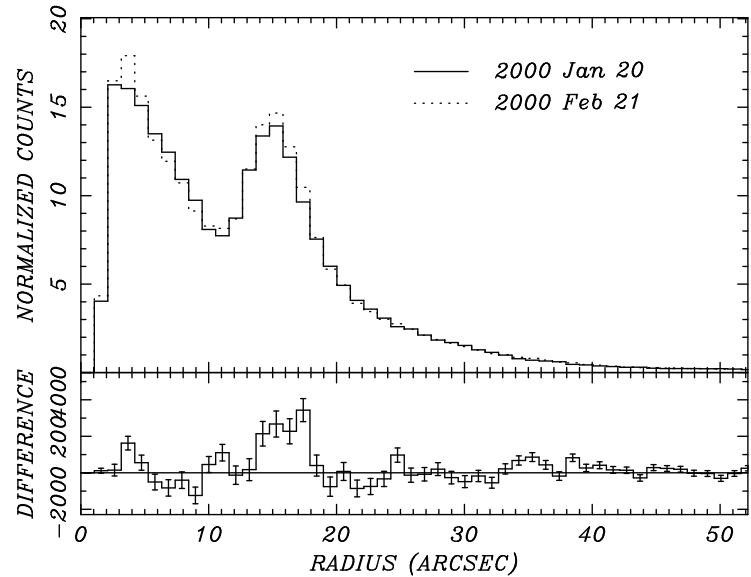}
}
\caption{Top panel: The radial profile of the Vela wind nebula
in elliptical annuli. The profile is centered on the pulsar which has been
suppressed and is restricted to the azimuthal sector delineated in Figure 4.
Bottom panel: The difference between the radial profiles for the two 
observations. Note the $7.8 \sigma$ excess in the second observation between
 $13.5^{\prime\prime}$ and $18^{\prime\prime}$ from the pulsar.}
\end{figure}

\begin{figure}
\centerline{
\psfig{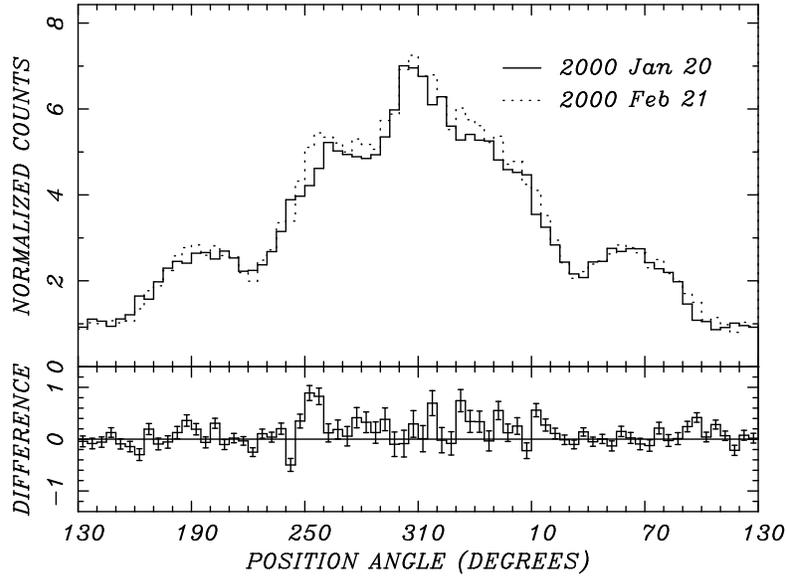}
}
\caption{The azimuthal distribution of surface brightness in an
elliptical annular ring between $13.5^{\prime\prime}$ and
$18^{\prime\prime}$ from the pulsar (upper panel).  The $7.6 \sigma$ excess
(see Fig 6) found in the second observation (lower panel) is located between position
angles $240^\circ$ and $10^\circ$. If caused by the glitch, this excess 
requires signal propogation at $0.7c$.  
}
\end{figure}

\begin{figure}
\centerline{ {\hfil\hfil
\psfig{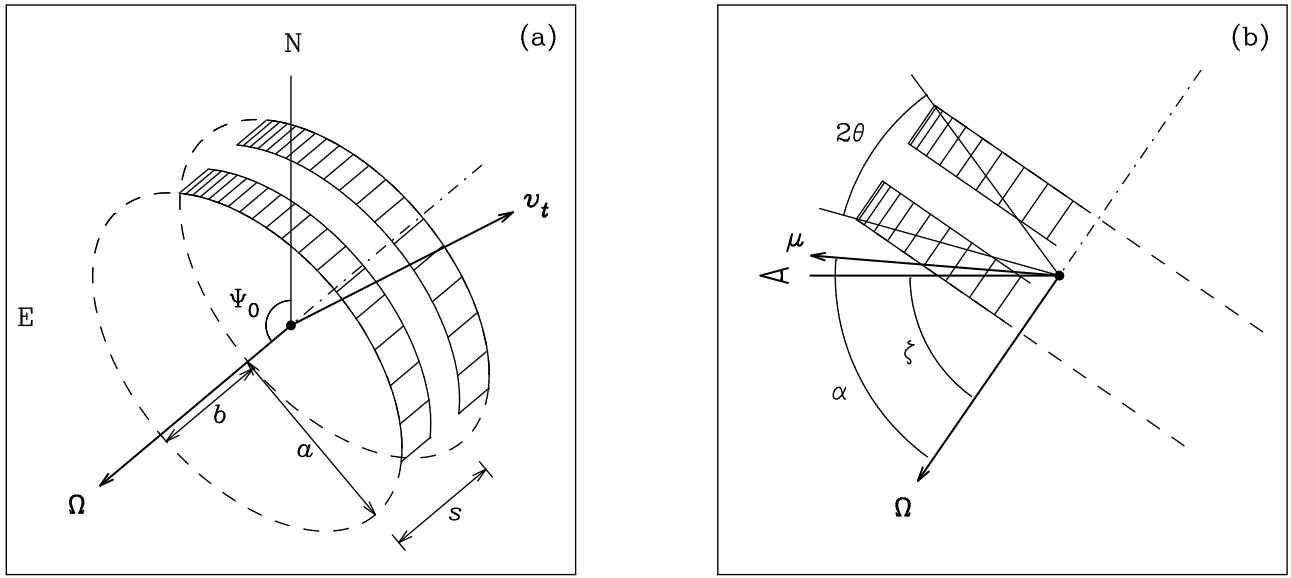}
\hfil\hfil} }
\caption{a) An idealized geometry for the Vela X-ray nebula as seen
in the {\it Chandra} images.  The two bright arcs are assumed to lie
on the front surface of an equatorial, toroidal wind whose axis of
symmetry is the rotation axis $\Omega$ of the neutron star. Lengths
$a$ and $b$ are the semi-major and semi-minor axes of the projected
ellipses, and $s$ is the projected distance between them.  For
scaling, $a = 25.7^{\prime\prime}$ translates to $0.96 \times
10^{17}$~cm at an assumed distance of 250~pc.  The vector $v_t$ is the
direction of proper motion of the pulsar.  b) A side view showing the
definitions of the angles between the rotation axis $\Omega$, the
magnetic axis $\mu$, and the line of sight.  The angle $\alpha -
\zeta$ is inferred from radio pulse polarization measurements, and
$\zeta$ comes from deprojection of the X-ray structures.  The angle
$\alpha$ is only approximately known.
}
\end{figure}

\end{document}